# Introducing an innovative robot-based mobile platform for programming learning

Konstantinos Manousaridis[1], Apostolos Mavridis[2], Konstantinos Anagnostopoulos[1], Gregory Kalogiannis[3]

[1]Computer and Mathematics Department
Mediterranean College of Thessaloniki
Thessaloniki, Greece

[2]Department of Informatics
Aristotle University of Thessaloniki
Thessaloniki, Greece

[3]School of Electrical and Computer Engineering
Aristotle University of Thessaloniki
Thessaloniki, Greece

***Abstract*—The present study introduces an Android based application that focuses on promoting students' learning experience when interacting with Lego Mindstorms Robotic Kit NXT. A thorough investigation of the current literature reveals that there are numerous applications attempting to overcome the limitations of Lego platform which, however, appear to only partially succeed in significantly improving children's engagement in fruitful learning. Their main limitations are that they either function on an elemental level or suffer from a general lack of creating effectively room for children to test their abilities and other curiosities. Aiming at confronting these limitations, the proposed android application, which is referred to as MyNXT, was created and presented. MyNXT offers direct access control along with the choice to program using simple linear logic through an easy and a user friendly environment with built in instructions for all ages. Concluding, the proposed application constitutes a stepping to an educating platform, which can facilitate the learning needs of children in contemporary times.**

## I. Introduction

This study researches the importance of current programs, application platforms and work done around Lego Mindstorms. Lego is a toy manufacturing company for children. Their products – toys involve and promote children reasoning, creativity and invoke the architects mind as the child "melds" the parts into many different structures. The present study targets Lego's intelligent brick NXT, emphasizing on the interaction and behavior.

Several studies prove children's robot-aided learning to be extremely beneficial for them with a positive impact on their overall behavior, self-motivation and learning ability [2, 3, 4, 6, 7, 8, 10]. These benefits are believed to be even greater when children are given an insight to the robot's logic and participate in its programming; one such case is the LEGO Robotic Kit Mindstorms.

Furthermore, the work at hand investigates students' learning experience and behavior when interacting with the Lego Mindstorms and possible ways of improving it through the deployment of an easy-to-use android application. The fact that current approaches assume the guidance of an instructor or prior knowledge of programming renders them rather less effective to specific group of ages, especially in earlier ages. Subsequently, there is a greater need for an application aimed at children that will increase their learning capabilities towards cognitive thinking and logic programming along with mobilizing a more enthusiastic engagement.

The proposed solution and final objective of this study is the creation of an android application that will facilitate students' learning experience when interacting with Lego Mindstorms Robotic Kit NXT. The application created will be referred to as MyNXT and was developed using MIT's app inventor online platform for designing android applications [1].

The paper is structured as follows: the first section presents a literature review focused on the use of Lego Mindstorms in education. In section 3, the methodology used in this study is analyzed while the final section summarizes the conclusions.

## II. Literature Review

Robots are mechanical or virtual artificial agents, usually an electro-mechanical machine that is guided by a computer program or electronic circuit to be autonomous or semi - autonomous; they can also be productively used as educational tools, from kindergarten to university contexts in many parts of the world [2]. Their usefulness is not limited in the industrial arena, but in others, as well, with an emphasis to the field of education. The latter is attributed to the fact that numerous studies have shown that when robots are used as tools for learning, they can facilitate the development of abstract thinking and collaborative problem solving as well as supporting learning in the various specific scientific, literary and artistic disciplines prescribed by standard school curricula [3, 4, 5, 6, 7, 8, 9, 10].

As far as teachers are concerned, many consider robots-based teaching to be highly motivating for students [11]. In relation to this, LEGO robots have been used in classrooms at various grade levels -including college level education- for the past 20 years [12]. LEGO Mindstorms are educational robotic systems with tailor-made hardware, software and educational resources that are used in classrooms, after-school club environments, and home schooling [13]. Through Mindstorms, students learn how to design, program and control fully-functional models and robots that carry out life-like automated tasks [13]. In what follows we present a discussion on extant literature pertaining to Mindstorms, and how it can (or cannot) enhance behavior and interaction among students in the classroom.

### A. *Effective use of Lego Mindstorms in education*

Koller & Kruijff [14] reported study results showing that in a computational linguistics course, students had been able to create simple yet interesting talking robots through the use of Mindstorms in just seven weeks. Kay [15] notes that using robotics lab exercises in an introductory course in robotics for undergraduates, with little or no experience in robot construction has shown encouraging results. Students demonstrated enhanced skills in software development, artificial intelligence and algorithmic functions, instead of low-level hardware control [15].

Wolff and Wahde [16] also state that another significant aspect of using robotics as a pedagogical tool pertain to the broad range of possibilities for the integration of classical engineering subjects, such as mechanics, electronics, software development, control theory and machine vision, with topics that mainly focus on psychology and cognition, such as human-robot interactions, within one interdisciplinary curriculum. However, it must be emphasized here that robotics should not be used to replace courses in control theory or machine vision but may be used as an opportunity to apply concepts learned from these courses.

Karp et al., [17] describe a GEAR program for elementary schools where teams, involving teachers and freshman students, were able to use robotics to improve the motivation of engineering in children while they learn concepts about electronics and computers. Garcıa-Cerezo, de Gabriel, Fernandez-Lozano, Mandow, Noz, Vidal-Verdu and Janschek [18] report the results of the 2nd Summer School on Mechatronics in an intercultural program between the University of Malaga, Spain and the University of Dresden, Germany. The researchers note that through the use of robotics, the participating students had been able to gain theory and practice competences on fuzzy logic, CAD design, Visual Programming, and real-time applications. Robotics, as mentioned earlier, is also used in engineering higher education.

There are many other studies that have been conducted on the effectiveness of robotic systems as pedagogical tools. The majority of these studies use project-based learning methodology to put basic robotic theory concepts in practice. Tools used in connection/combination with these robotics courses or programs, range from standard off-the-shelf robotic kits such as Mindstorms and the Boe-Bot from Parallax Inc., to more specialized and custom-built equipment, such as a bipedal humanoid robot or a research-grade mobile robot platform [16]. On student's side, these projects provide additional benefits in the sense that they become highly motivated to acquire further knowledge and enhance their creativity.

### B. *Lego Mindstorms Programming environments*

There is also a broad range of standard programming environments for the LEGO NXT brick. Although there have been numerous Mindstorms model designs created by experts through the use of sophisticated programs [19], there are more recent ones, such as those described in the following paragraphs, that are appropriate for students.

NXT-G is the software included in the primary commercial Mindstorms' pack for educational purposes. It is a visual programming environment based on block diagrams. The user can choose blocks and join them, so that the desired robot behavior is achieved. This is most suitable for children because the program is easy to use [20]. Another programming environment is the Microsoft Robotic Studio (MSRS), which is applicable for all robotics. It supports block diagrams and drag-and-drop features for the programming using language MVPL. Although this software is flexible, it is not as easy to use as NXT-G.

The RWTH-Mindstorms NXT toolbox is a Matlab toolbox developed at the RWTH Aachen University [20]. It harnesses the computational power of the Matlab environment with a command line function to control the NXT brick and is thus supported through Windows and Linux. Within this environment, the PC processor performs the main workload. Complex engineering tasks including computer vision or optimization algorithms may be implemented. This is a sophisticated program which is appropriate only for early-course engineering students [20]. On the other hand, the NeXT Byte Codes (NBC) is an assembler-like language that allows low-level NXT programming [20]. It provides users with a thorough instruction set including, subroutines, macros and compiler pre-processing commands, similar to C. It is an open source solution and is supported under Windows and Linux platforms. Due to the complexity of the language used in this program, it is most suitable for undergraduate students in Electrical or Computer Engineering courses focusing on embedded systems or computer architecture [20]. Another program is Not eXactly C (NXC), which is similar to C [20]. It is supported through NBC, and "*provides the flexibility of this language with the structure of a middle-level language such as C*" [20]. It is open source and supported from both Windows and Linux. There is another programming environment for NXC under Windows, called Bricx Command Center [20]. This program has a text editor, NXC compiler, and software to send the programs to the NXT brick. This language is appropriate for computer engineering students because of its flexibility and complexity.

In light of the broad spectrum and varying complexities of the languages used by the NXT brick, Mindstorms has been used by instructors throughout primary, middle, high school, undergraduate or even graduate engineering projects. A distinct advantage of Mindstorms is the flexibility of its platform. Learners are able to build robots through the use of the LEGO classic pieces such that they can achieve their project requirements. They also have a relatively high degree of freedom in using sensors and motors, building complex structures and so forth. The platform may be used and subsequently reused in distinct applications and experiments while maintaining the same laboratory cost [20].

### C. *Robot controlling applications*

There are different approaches on how to control a robot or in this particular case, LEGO Mindstorms through an application. Studies that introduce and utilize applications to control robots both for creating a mobile robot application infrastructure as well as studying the educational effect they have on students exist. Most of which create a Bluetooth connection between the device and the robot in order to achieve communication. Such approaches so far enable the user to directly control the robot through voice or tactile commands [26]. However, there is a need for an application that allows the user to program commands and forward them to the robot.

Programming or controlling a robot for educational purposes is better facilitated through a mobile application than a desktop program [26]. Mobile devices offer portability thus enabling the user to function in any environment. Most students are already familiar with smartphones and applications nowadays, eliminating the need to teach the basics of a device. Last but not least mobile devices tend to be cheaper than desktop computers.

## III. Methodology

Although Lego Mindstorms and the NXT platform are well documented and researched on, there are a number of limitations that makes it very difficult for children to use it in the absence of an instructor or prior programming knowledge.

### A. *Difficulties in Lego Programming*

Foremost, largely Lego programming is not code programming, but a command box programming. Just like other Command Line Interface (CLI), Smutny [26] acknowledges that command box programming demands that the users (and children, in this case) have prior programming knowledge for it to function effectively. This relates the effectiveness of Lego NXT in direct proportion to the level of knowledge of commands that the users are aware of [21]. Furthermore, the aforementioned limitation also extends to most Lego-supplied languages, namely RCX Code, NXT-G Code, ROBOLAB (LabVIEW and a C-Based Programming Language), and Microsoft Robotics. Legos' development platform has other severe limitations and a big blunder is the fact that it lacks direct control. Overall speaking, Lego limitedly supports the creation of interactive projects, which are capable of communicating with the users and the computer device while the project is running [22].

Commenting further on the complexity of Lego, Lew, Horton, and Sherriff [23] demonstrated that even to students that are familiar with software engineering principles, Lego NXT still proved challenging enough, and it was being used as a basis for a complex, communicating system. Lego Mindstorms NXT is a highly beneficial set for teaching advanced software development, but of limited usefulness regarding children. It has been used widely for introductory purposes in most computer science courses. However, it is reported that Lego platform has not been extensively applied in simulated production environment [23], which would be highly beneficial to children because it needs implementation of command protocols capable of increasing their cognitive thinking and logic programming skills, plus this would engage them more enthusiastically [2, 3, 4, 6, 7, 8, 10].

The limitations of Lego platform are quite evident despite the highly propagated notion that Lego is suitable for most children in terms of professional curiosity, cost, child interest and above all and most relevant to our situation, flexibility. However, Lego kits and specifically Mindstorms lack compatibility to the most commonly used program languages C/C++ and Java. The reality forces instructors and educators to first familiarize children or students to the main features of new and traditional programming languages, before they can make use of the Lego platform [22]. The elementary platform lacks support for students to enable them pursue floating-point arithmetic, while the limitation is not critical, it avails serious challenges to children and students who want to introduce themselves to mathematics that support robotic navigation.

To give a vivid background, Cuevas, Zaldivar, and Pérez-Cisneros [22] reaffirms that “LEGO provides two tools for programming the Intelligent brick; the first is a development environment for programming the Intelligent brick with an interface that models programming as a process of dragging puzzle pieces (representing program steps) together to produce a chain (complete program)”. To improve the overall Lego capability and intelligent brick applicability, RIS (Robotics Invention System) was introduced. Arguably, RIS is depicted as simple enough for children; however, it has various limitations especially on robot state expression. Subsequently, numerous lobbyists have substituted RIS, which is a graphical programming language, with “standard” programming languages such as Java, Scheme, Ada, and C; hence, the introduction and popularization of ROBLAB, LabVIEW, legOS, leJOS, etc. However, all these modules, packages and toolkits make the implementation for common programming languages easier, especially in integrating them into the academic curriculum. Still, they only partially succeed in making Lego user-friendly and hence appropriate for children with no prior programming knowledge.

At this point, it is rather evident that the current programming languages, programs and work done around NXT are limiting. Notably, there are several attempts to solve the aforementioned limitations, but are still far from being truly effective. In a philosophical point of view, Chau [24] highlights that most mobile apps under the NXT platform fail to accommodate appropriately the intellectual ability and developmental stages of children.

### B. *Current solutions in Lego Mindstorms programming platforms*

One of the most recognized groups that exposed the limitations of the Lego platform and subsequently developed their package of tools -named as MTM- was Cuevas, Zaldivar, and Pérez-Cisneros [22]. They first described the limitations of Lego Mindstorms robots, in both software and hardware, and then analyzed certain third-party programming environments capable of overcoming the identified limitations. In a last step, they outlined the means through which their own work would eliminate most of the remaining limitations. Notably, Cuevas, Zaldivar, and Pérez-Cisneros [22] provide a very vivid background because their MTM availed a firmware that permits point-to-point communication, C++ API development kits focuses on programming the robot agents while reading the robot status. In addition, MTM includes software packages in Java and Lisp for overall robotic programming while aiming in remotely controlling them.

The MTM Tool Set attempted with a notable success to improve the applicability of Lego and render it simple for both instructors and children. In sum, it is an enhanced version of Lego Mindstorms original software that can now support a number of extra functions. Moreover, the most closely related to the subject under discussion is the fact that the native-language compiler, which permits the children to execute on-the-fly compiling, and, above all, download the code to the Lego intelligent brick. In their conclusion, Cuevas, Zaldivar, and Pérez-Cisneros [22] emphasized that there is a need for more apps and other elements that can further enhance the simplicity and usability of Lego to children.

Bowers [25] takes a different look at the most prominent children apps in order to demonstrate the simplicity needed in an Android application intended for teaching children and increasing their learning capabilities towards cognitive thinking and logic programming along with engaging them more enthusiastically. First and foremost, Bowers [25] observes that the best apps for children are cantered more on simple animation, graphics, and actual codes, in general. Such educational apps are capable of demonstrating the understanding of cause and effect, and further motivating the children to learn more about programming. The most common and simplest of these apps is the “Daisy the Dinosaur” (for iPad, and available for free). The app prompts the children to manipulate a character, named Daisy. The manipulations are being attained through a number of challenges, involving events, loops, and other programming fundamental concepts. For example, when the children touch the dinosaur, it moves forward. Despite its simplicity, it greatly helps students in terms of developing their cognitive ability and logic programming skills [25].

The other app that Bowers [25] highlights is “Move the Turtle” which involves basic logical programming conceptions through the manipulation of a single graphical object and entails a number of challenges. Ironically, the cute turtle is comparable to the Lego depictions back in the 60s all through to the 80s; however, it serves the purpose more than the current versions of Lego. Remarkably, both “Daisy the Dinosaur” and “Move the Turtle” make the children learn a great deal towards the aforementioned skills and concepts.

### *C. The proposed solution*

By taking into consideration all the above research, one comes to the conclusion that there is a need to provide an easy to use platform incorporating the use of a simple robotic kit; this will grant children the ability to build, learn and communicate with the robot. The proposed application offers direct control over the motors control by the intelligent brick NXT and basic linear programming.

Upon launching the application on any android device, the first screen, i.e., the home screen appears (Fig. 1). There are five buttons in the first screen. The first three buttons open the corresponding screens, which offer the user three types of direct control over Bluetooth to the NXT. The first type, speech recognition, recognizes specified voice commands on the touch of the microphone icon. Those commands are given to the user at the bottom of the screen. The second type, accelerometer, takes advantage of the android device’s build in accelerometer to transfer and translate the movements of the device to moving the robot forward, backward, left and right. Last, the third type, arrow keys, is a common button control. The fourth button, name Logic Creator, opens the Logic Creator screen as seen on the right half of Fig. 1, which allows the user to prebuild the logic that the robot will later on execute when the run button is pressed. Hence, children are allowed to create and execute their own linear logic. The functionality of the Logic Creator is depicted in Fig. 2.

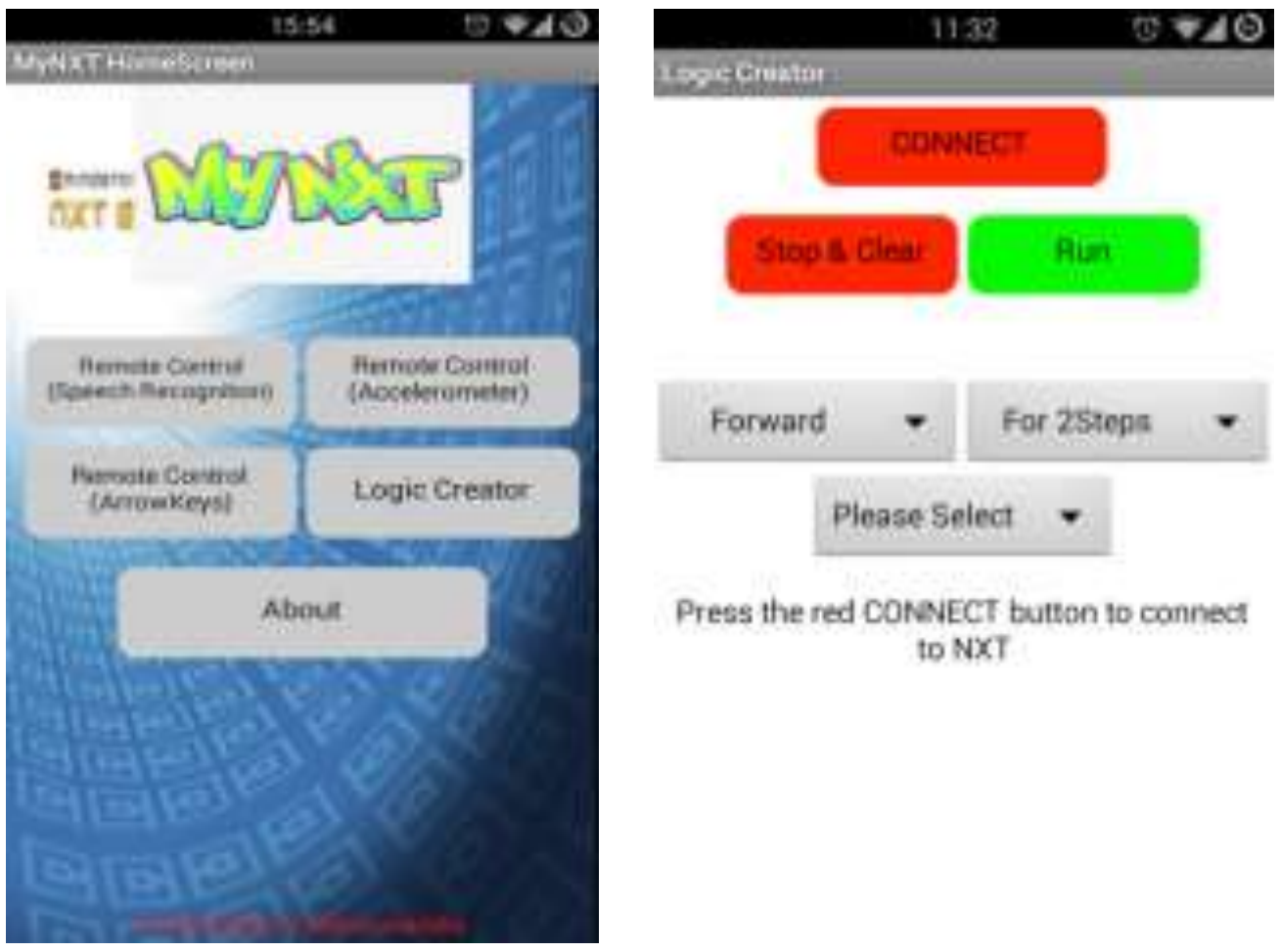

(a) The home screen of MyNXT (b) The Logic Creator of MyNXT

Fig. 1. MyNXT’s home screen (a) and the Logic Creator function (b)

In order for the application to be simple and engaging for children, simple instructions are given at all stages. In every control screen there is an indicating label that guides the user to the proper function of the application. As depicted in Fig. 1, when entering any control screen the label prompts the user to press the “CONNECT” button in order to connect the device with the robot. After the connection is established the label in each control screen instructs the user with the next step in order to control the robot. For example in the “Speech Recognition” control screen the indicator reads “Press the Microphone to Give Commands”. Programming of the app has also taken into account defensive design in order to assure that there will always be written instruction within the application in order to avoid potential misuse or mishap. For example, if the Bluetooth on the device is turned off when the “CONNECT” button is pressed, a pop-up warning is issued, informing the user to turn the Bluetooth on. The app, by exploiting the direct control as a stepping-stone, achieves early engagement of younger children by slowly directing them to the logic creator which can later be extensively used for learning through various quests or task.

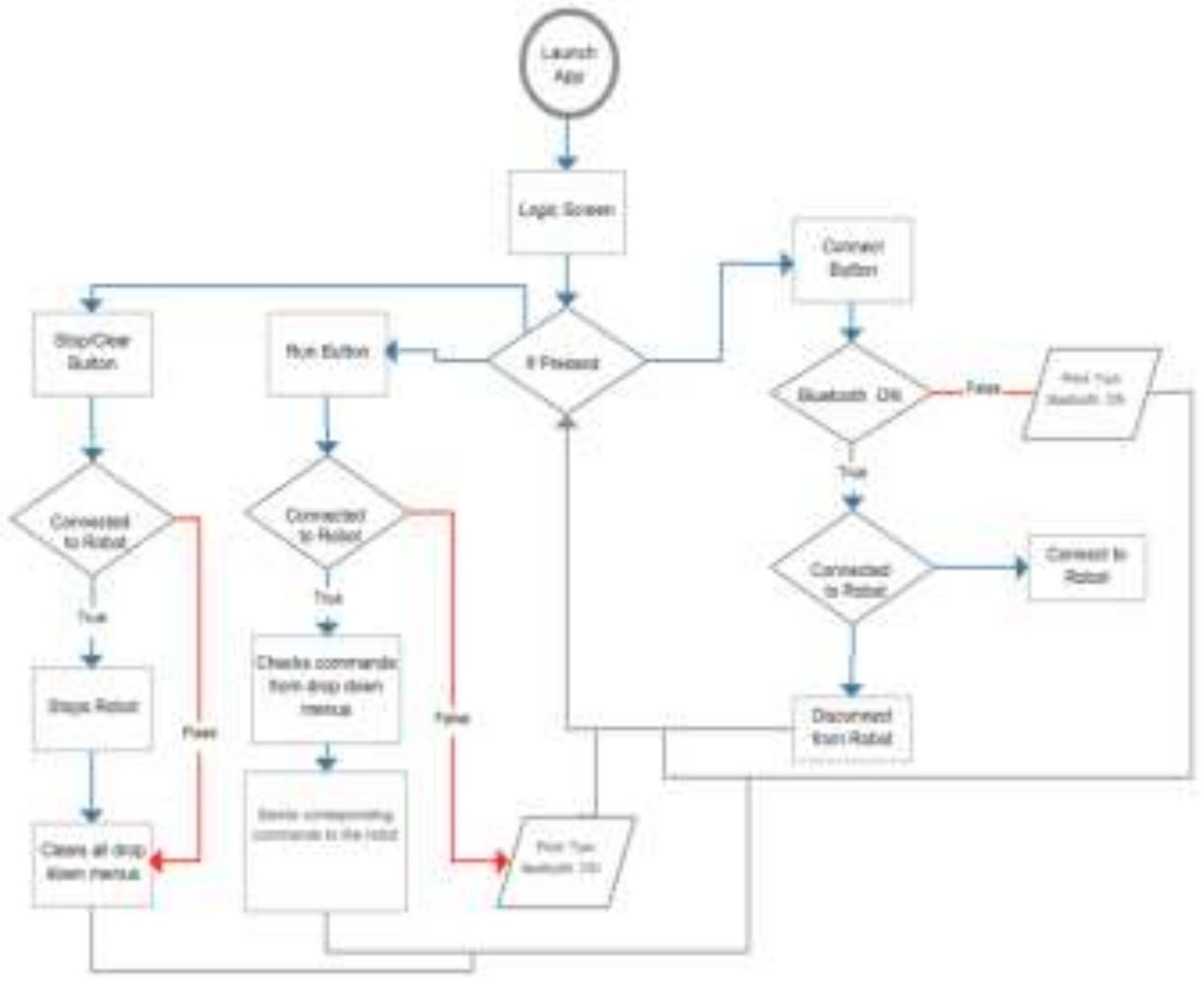

Fig. 2. Flowchart of the Logic Creator

## IV. CONCLUSION

Lego platform is extremely complicated for children and it is not attaining the desired effect. The current programming languages, programs, and research work concerning NXT are limited, due to their complexity, to be applied with the assistance of an instructor or with the assumption of prior programming knowledge. The main implication is the fact that there is a greater need for more applications, which effectively and appropriately engage children and their learning capabilities towards cognitive thinking and logic programming. Finally, there is also a necessity to engage other senses and interactions, such as the use of audio instructions because the touch manipulation has been over exploited.

MyNXT application mainly targets students of younger ages, 8 to 15 years old. The fast evolution of Computer Science creates a need for earlier engagement towards computer programming. MyNXT fills that need utilizing the Lego Mindstorms, which is mostly considered by children a game, engaging them to such direction. Especially taking into consideration the fact that children nowadays have interaction with mobile devices from early ages.

MyNXT is based on a modular architecture that makes it easy to extend the main application by adding more features. There is actually no limit to the extensions that can be developed, given enough time. For instance, there could be an implementation for the various Lego sensors to provide the user quite more options and flexibility. Another approach could be to create a quest line with specific tasks that should be finalized by the user in order to correctly complete the quest with the minimum commands possible. Furthermore, different approaches can also be implemented to the user interface interaction of the Logic Creator screen, such as voice commands. Additionally many of the in-app text messages and warnings can be implemented to be alternative provided as audio messages.

## REFERENCES

[1] Appinventor.mit.edu, 'MIT App Inventor | Explore MIT App Inventor', 2015. [Online]. Available: http://appinventor.mit.edu/explore/. [Accessed: 02- Feb- 2015].

[2] E. Datteri, L. Zecca, F. Laudisa and M. Castiglioni, 'Learning to explain: the role of educational robots in science education', Themes in Science and Technology Education, vol. 6, no. 1, pp. pp. 29-38, 2013.

[3] M. Bers and M. Portsmore, 'Teaching Partnerships: Early Childhood and Engineering Students Teaching Math and Science Through Robotics', Journal of Science Education and Technology, vol. 14, no. 1, pp. 59-73, 2005.

[4] M. Petre and B. Price, 'Using Robotics to Motivate "Back Door" Learning', Education and Information Technologies, vol. 9, no. 2, pp. 147-158, 2004.

[5] J. Murray and K. Bartelmay, 'Inventors in the Making.', Science and Children, vol. 42, no. 4, pp. 40-44, 2005.

[6] M. Barak and Y. Zadok, 'Robotics projects and learning concepts in science, technology and problem solving', Int J Technol Des Educ, vol. 19, no. 3, pp. 289-307, 2007.

[7] G. Ardito, P. Mosley and L. Scollins, 'WE, ROBOT: Using Robotics to Promote Collaborative and Mathematics Learning in a Middle School Classroom', Middle Grades Research Journal, vol. 9, no. 3, p. 73, 2014.

[8] A. Eguchi, 'Educational robotics for promoting 21st century skills', Journal of Automation Mobile Robotics and Intelligent Systems, no. 8, 1, pp. 5--11, 2014.

[9] W. Hwang and S. Wu, 'A case study of collaboration with multi-robots and its effect on children's interaction', Interactive Learning Environments, vol. 22, no. 4, pp. 429-443, 2012.

[10] T. Yuen, M. Boecking, J. Stone, E. Tiger, A. Gomez, A. Guillen and A. Arreguin, 'Group Tasks, Activities, Dynamics, and Interactions in Collaborative Robotics Projects with Elementary and Middle School Children', Journal of STEM Education: Innovations and Research, vol. 15, no. 1, 2014.

[11] A. Hirst, J. Johnson, M. Petre, B. Price and M. Richards, 'What is the best programming environment/language for teaching robotics using Lego Mindstorms?', Artif Life Robotics, vol. 7, no. 3, pp. 124-131, 2003.

[12] E. Danahy, E. Wang, J. Brockman, A. Carberry, B. Shapiro and C. B., 'LEGO-based Robotics in Higher Education: 15 Years of Student Creativity', Int J Adv Robotic Sy, p. 1, 2014.

[13] C. Mellon, 'Lego Mindstorms', Education.rec.ri.cmu.edu, 2014. [Online]. Available: http://www.education.rec.ri.cmu.edu/content/lego/start/. [Accessed: 10- Jan- 2015].

[14] A. Koller and G. Kruijff, 'Talking robots with LEGO MindStorms', Proceedings of the 20th international conference on Computational Linguistics - COLING '04, 2004.

[15] J. Kay, 'Two Lab Exercises for an Introductory Robotics Class', Aaai.org, 2004. [Online]. Available: http://www.aaai.org/Library/Symposia/Spring/2004/ss04-01-033.php. [Accessed: 03- Jan- 2015].

[16] K. Wolff and M. Wahde, 'Balancing Theory and Practical Work in a Humanoid Robotics Course.', International Journal of Teaching and Learning in Higher Education, vol. 22, no. 1, pp. 80-88, 2010.

[17] T. Karp, R. Gale, L. Lowe, V. Medina and E. Beutlich, 'Generation NXT: Building Young Engineers With LEGOs', IEEE Trans. Educ., vol. 53, no. 1, pp. 80-87, 2010.

[18] A. Garcia-Cerezo, J. Gomez-de-Gabriel, J. Fernandez-Lozano, A. Mandow, V. Munoz, F. Vidal-Verdu and K. Janschek, 'Using LEGO robots with LabVIEW for a Summer School on Mechatronics', 2009 IEEE International Conference on Mechatronics, 2009.

[19] M. Agullo, LEGO Mindstorm masterpieces. Rockland, Mass.: Syngress, 2003.

[20] M. CuÃ©llar and M. Pegalajar, 'Design and implementation of intelligent systems with LEGO Mindstorms for undergraduate computer engineers', Comput. Appl. Eng. Educ., vol. 22, no. 1, pp. 153-166, 2011.

[21] E. Klopfer, S. Osterweil and K. Salen, 'Moving learning games forward', Halshs.archives-ouvertes.fr, 2009. [Online]. Available:

https://halshs.archives-ouvertes.fr/hal-00593085/. [Accessed: 02- May-2015].

[22] E. Cuevas, D. Zaldivar and M. Pérez-Cisneros, 'Low-Cost Commercial Lego Platform for Mobile Robotics', International Journal of Electrical Engineering Education, vol. 47, no. 2, pp. 132-150, 2015.

[23] M. Lew, T. Horton and M. Sherriff, 'Using LEGO MINDSTORMS NXT and LEJOS in an Advanced Software Engineering Course', 2010 23rd IEEE Conference on Software Engineering Education and Training, 2010.

[24] C. Chau, 'Positive Technological Development for Young Children in the Context of Children's Mobile Apps', PhD, Tufts University, 2014.

[25] B. Bowers, 'A look at early childhood programming in museums.', Journal of Museum Education, vol. 37, no. 1, pp. 39-48, 2012.

[26] J. Nadvornik and P. Smutny, 'Remote control robot using Android mobile device', Proceedings of the 2014 15th International Carpathian Control Conference (ICCC), 2014.